\documentclass[runningheads]{llncs}

\usepackage{eccv}
\usepackage{eccvabbrv}

\usepackage{amsmath,amssymb}
\usepackage{booktabs}
\usepackage{graphicx}
\usepackage{microtype}
\usepackage[section]{placeins}
\usepackage{xcolor}
\usepackage{colortbl}
\usepackage{tikz}
\usetikzlibrary{arrows.meta,fit,positioning,shapes.geometric}
\usepackage[accsupp]{axessibility}
\usepackage{hyperref}
\usepackage{orcidlink}
\hypersetup{hidelinks}

\definecolor{ink}{HTML}{3D4853}
\definecolor{softgray}{HTML}{F3F5F7}
\definecolor{accentblue}{HTML}{3178B8}
\definecolor{accentgreen}{HTML}{4E9C6A}
\definecolor{accentorange}{HTML}{C77B39}
\definecolor{accentpurple}{HTML}{8065A8}
\definecolor{bluewash}{HTML}{E8F2FF}
\definecolor{greenwash}{HTML}{E2F3E8}
\definecolor{orangewash}{HTML}{FFF0E3}
\definecolor{purplewash}{HTML}{EEE7F7}
\definecolor{yellowwash}{HTML}{FFF5CC}

\newcommand{\openaiicon}{\leavevmode\hbox to 1.35em{%
  \pdfliteral{q 0 0 0 rg 11.5 0 0 11.5 0 -1.8 cm
  BI /W 64 /H 64 /IM true /BPC 1 /F /AHx /D [0 1] ID
  0000000000000000000000000000000000000000000000000000000000000000000000FE00000000000003FFC000000000000FFFF000000000001FFFF8F00000
  00007E00FFFF0000000078003FFF80000000F000FFFFE0000001E001FC07F0000001E007F000F8000001C01FC00078000003C03F00003C00000780FE00001E00
  001F83F801C01E00007F83E007F00E0000FF83C01FF80F0001FB83803FFE0F0001E383C0FE3F8F0003C383C3F80FEF00078383CFF007FF000783839FF801FF00
  070383FFFE007E000F0383FC3F801F000F0383F00FC00F800E0383C003F007C00E03838001FC03C00E0383C003FE01E00F0383C003DF80E00F03C3C003CFC0F0
  0F03F3C003C3C0F00701FFC003C1C0F007807F8001C1C0F003C03FC003C1C07003E00FE007C1C0F001F003F81FC1C0F000FC00FC3FC1C0F000FE007FFFC1C0E0
  00FF801FF9C1C1E000FFE00FE3C1C3E000F3F83F83C1C3C000F0FC7F03C1CF8000F07FFC01C1FF0000F01FF003C1FE00007007C00FC1FC00007803801F81F800
  007800007F01C000003C0001FC03C000001E0003F0038000001F800FC0078000000FE03F800700000003FFFE000F00000001FFFC003E000000007FFF00FC0000
  0000001FFFF8000000000007FFF0000000000001FFC00000000000007E0000000000000000000000000000000000000000000000000000000000000000000000>
  EI Q}\hss\vrule width0pt height1.05em depth0.18em}}

\newcommand{\qwenicon}{\leavevmode\hbox to 1.35em{%
  \pdfliteral{q 0.38 0.36 0.93 rg 11.5 0 0 11.5 0 -1.8 cm
  BI /W 64 /H 64 /IM true /BPC 1 /F /AHx /D [0 1] ID
  0000000000000000000000000000000000000000000000000000000000000000000001FF80000000000001FFC0000000000003FFE00000000000033FF0000000
  0000073FF000000000000E1FF800000000000E1FFFFFF00000001C0FFFFFF8000000180FFFFFF8000000380FFFFFFC000000301FFFFFFE000000703FFFFFFE00
  0000E03FFFFFFF0000FFE07FFFFFFF0000FFC07FFFFFFF0001C000000000030003E000000000070003E000000000070007F0000000000E0007F8000000000C00
  0FF8000000001C000FFC0000000018000FFC01FFFF801C000FFE01FFFF800C0007FF00FFFF000E0007FF00FFFE00060003FF807FFE00070003FF803FFC000380
  01FFC03FFC01038000FFC01FFC0301C000FFE01FF80780C0007FF00FF007C0E0007FF00FF00FC060003FF807E00FE070003FF803E01FFFF0001FFC03C01FFFF0
  003FFC01803FFFF0007FFE01807FFFE0007FFE00007FFFE000FFFF0000FFFFC000FFFF8000FFFFC000FFFF0001FFFF8000FF800001FFFF0000FF000003FF8000
  007E000007FF0000007E000007FE0000003C00000FFE0000001C00000FFE0000001FFFF81FFC00000007FFFC1FF800000000000C3FF000000000000E7FF00000
  000000077FF0000000000007FFE0000000000003FFE0000000000001FFC000000000000000000000000000000000000000000000000000000000000000000000>
  EI Q}\hss\vrule width0pt height1.05em depth0.18em}}

\newcommand{\JF}{\mathcal{J}\&\mathcal{F}}
\newcommand{\Nacc}{\mathrm{N\mbox{-}acc.}}
\newcommand{\Tacc}{\mathrm{T\mbox{-}acc.}}
\newcommand{\Final}{\mathrm{Final}}
\newcommand{\method}{\textsc{Speech2MaskTrack}}

\begin{document}

\title{Motion-Aware Reasoning from Speech to Mask Tracks:\
Runner-up Solution for the MeViS-Audio Track of the 8th LSVOS Challenge 2026}
\titlerunning{Motion-Aware Reasoning from Speech to Mask Tracks}

\author{Jinxing Zhou\inst{1}\orcidlink{0000-0001-6402-7593} \and
Suiyi Zhao\inst{2} \and
Yanghao Zhou\inst{3} \and
Ruohao Guo\inst{4}\orcidlink{0000-0002-1091-272X}}
\authorrunning{J. Zhou et al.}

\institute{Mohamed bin Zayed University of Artificial Intelligence \and
Anhui University of Science and Technology
 \and National University of Singapore\and
China Agricultural University}

\maketitle

\begin{abstract}
Speech-guided referring video object segmentation aims to recover the mask
tracks of objects specified by a spoken motion description. Here, speech
carries a linguistic instruction rather than acoustic evidence from a sounding
object, so a solution must connect speech recognition, motion-centric temporal
grounding, mask tracking, and explicit no-target handling. We introduce
\method, our approach for the MeViS-Audio track of the 8th LSVOS Challenge.
\method{} transcribes the spoken query and
compiles it into structured constraints over category, count, direction,
interaction role, and temporal phase. SAM3.1 enumerates multiple instance
tracks, which TRACE ranks using complete-trajectory motion and relation
evidence. A frozen lexical presence gate may suppress the ranked SAM3.1 base
prediction.
When the gate predicts that a target is present, an available
full-expression-conditioned SaSaSa2VA track replaces the SAM3.1 mask. Only
outputs that remain empty enter GPT-assisted recovery, which invokes
SaSaSa2VA again under query- and mask-level verification. \method{} achieved
second place in the official challenge ranking.
\keywords{Referring video object segmentation \and Speech-guided segmentation
\and Motion reasoning \and Video object segmentation}
\end{abstract}

\section{Introduction}
\label{sec:introduction}

The Large-scale Video Object Segmentation (LSVOS) Challenge is a long-running
international benchmark series for video object segmentation
(VOS)~\cite{MOSE,ReferringSurvey}. Since its origin in the large-scale
YouTube-VOS benchmark~\cite{xu2018youtube}, LSVOS has progressively moved from
relatively controlled videos toward long-term, crowded, occluded, and
motion-intensive real-world scenes~\cite{MOSEv2,Ding2023MeViSAL}. Its sixth
edition attracted 129 registered teams from more than 20 institutions across
over eight countries~\cite{Ding2024LSVOSCR}, and the seventh introduced MOSEv2
to further emphasize robust segmentation under realistic
conditions~\cite{MOSEv2,liu2025lsvos}. LSVOS therefore provides a widely used
community testbed for measuring progress and exposing failure modes hidden by
easier VOS benchmarks.

\begin{figure}[t]
  \centering
  \definecolor{maskcue}{HTML}{C98518}
  \definecolor{textcue}{HTML}{7656B5}
  \resizebox{\textwidth}{!}{%
  \begin{tikzpicture}[
    font=\sffamily,
    trackrow/.style={rounded corners=7pt, draw=ink!35, line width=0.9pt,
                     fill=white, minimum width=12.8cm, minimum height=1.55cm},
    focusrow/.style={rounded corners=7pt, draw=accentblue, line width=1.7pt,
                     fill=bluewash, minimum width=12.8cm, minimum height=1.55cm},
    input/.style={rounded corners=3pt, draw=ink!45, fill=softgray,
                  minimum width=2.55cm, minimum height=0.43cm,
                  align=center, font=\sffamily\scriptsize},
    cue/.style={input, minimum width=3.35cm, text width=3.05cm,
                inner xsep=0pt, align=center},
    task/.style={rounded corners=4pt, draw=ink!65, dashed, line width=1.0pt,
                 fill=white, minimum width=3.05cm, minimum height=0.86cm,
                 align=center, font=\sffamily\scriptsize},
    output/.style={rounded corners=4pt, draw=accentgreen!85!black,
                   fill=greenwash, minimum width=2.20cm,
                   minimum height=0.62cm, align=center,
                   font=\sffamily\scriptsize\bfseries},
    flow/.style={-{Latex[length=2.0mm]}, line width=1.05pt, draw=ink!70}
  ]
    \node[trackrow] (mose-row) at (6.40,0) {};
    \node[trackrow] (text-row) at (6.40,-2.05) {};
    \node[focusrow] (audio-row) at (6.40,-4.35) {};

    \node[font=\sffamily\bfseries\normalsize, text=ink, anchor=west]
          at (0.22,0.98) {Track 1: MOSEv2};
    \node[font=\sffamily\bfseries\normalsize, text=ink, anchor=west]
          at (0.22,-1.07) {Track 2: MeViSv2-Text};
    \node[font=\sffamily\bfseries\normalsize, text=accentblue, anchor=west]
          at (0.22,-3.37) {Track 3: MeViSv2-Audio};
    \node[rounded corners=4pt, fill=accentblue, text=white,
          font=\sffamily\scriptsize\bfseries, inner xsep=6pt, inner ysep=2pt,
          anchor=east] at (12.65,-3.37) {OUR FOCUS};

    \node[input] (mose-video) at (2.55,0.30) {Video frames};
    \node[cue, draw=maskcue, fill=maskcue!10] (mose-cue)
          at (2.55,-0.30) {First-frame target mask(s)};
    \node[input] (text-video) at (2.55,-1.75) {Video frames};
    \node[cue, draw=textcue, fill=textcue!10] (text-cue)
          at (2.55,-2.35) {Text motion expression};
    \node[input] (audio-video) at (2.55,-4.05) {Video frames};
    \node[cue, draw=accentblue, fill=accentblue!8] (audio-cue)
          at (2.55,-4.65) {Spoken motion expression};

    \node[task] (mose-task) at (7.05,0)
          {Semi-supervised VOS\\\textcolor{ink!70}{propagate initialized targets}};
    \node[task] (text-task) at (7.05,-2.05)
          {Referring VOS\\\textcolor{ink!70}{resolve a textual motion query}};
    \node[task, draw=accentblue, fill=bluewash] (audio-task) at (7.05,-4.35)
          {Speech-guided referring VOS\\\textcolor{ink!70}{resolve a spoken motion query}};

    \node[output] (mose-out) at (10.90,0) {Target\\mask tracks};
    \node[output] (text-out) at (10.90,-2.05) {Referred\\mask tracks};
    \node[output, draw=accentblue, fill=white] (audio-out) at (10.90,-4.35)
          {Referred\\mask tracks};

    \draw[flow] (mose-video.east) -- (mose-task.west |- mose-video.east);
    \draw[flow, draw=maskcue] (mose-cue.east)
          -- (mose-task.west |- mose-cue.east);
    \draw[flow] (mose-task.east) -- (mose-out.west);

    \draw[flow] (text-video.east) -- (text-task.west |- text-video.east);
    \draw[flow, draw=textcue] (text-cue.east)
          -- (text-task.west |- text-cue.east);
    \draw[flow] (text-task.east) -- (text-out.west);

    \draw[flow] (audio-video.east)
          -- (audio-task.west |- audio-video.east);
    \draw[flow, draw=accentblue] (audio-cue.east)
          -- (audio-task.west |- audio-cue.east);
    \draw[flow, draw=accentblue] (audio-task.east) -- (audio-out.west);
  \end{tikzpicture}}
  \caption{\textbf{The three tracks of the 8th LSVOS Challenge.} Each track combines
  a video with a different target specification and produces temporally
  consistent mask tracks. MOSEv2 initializes targets with first-frame masks,
  whereas MeViSv2-Text and MeViSv2-Audio identify targets from textual and
  spoken motion expressions, respectively. Our work focuses on MeViSv2-Audio;
  its audio is a spoken query rather than sound emitted by the target.}
  \label{fig:tracks}
\end{figure}
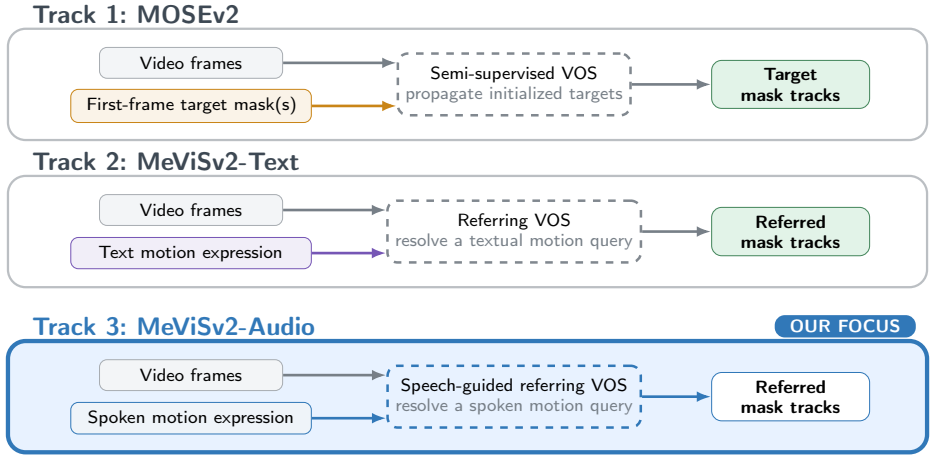

The 8th LSVOS Challenge, held in conjunction with ECCV 2026, contains three
complementary tracks, summarized in \cref{fig:tracks}. They differ primarily in
how target objects are specified. The MOSEv2 track adopts semi-supervised, or
one-shot, VOS: pixel-level masks in the first frame identify the target
instances, and a system must preserve their identities and predict their masks
throughout the remaining video. MOSEv2 stresses this propagation under severe
occlusion, disappearance and reappearance, crowded scenes, small objects,
adverse environments, and multi-shot transitions \cite{MOSEv2}. In contrast,
MeViSv2-Text and MeViSv2-Audio provide no first-frame target masks. A system
must identify the referred objects from textual or spoken motion expressions,
respectively, before segmenting them over time
\cite{lsvos2026}. The latter two tracks are based on MeViSv2~\cite{ding2025mevis}, a
multi-modal benchmark containing 33,072 human-annotated text and audio motion
expressions for 8,171 objects in 2,006 videos. Static appearance is often
insufficient: a target may be defined by its direction, interaction role,
temporal order, or behavior relative to visually similar instances. We focus
on the \textbf{MeViS-Audio} track, where these motion-centric referring expressions are
delivered as speech.

Despite the word \emph{audio}, MeViS-Audio differs fundamentally from
conventional audio-visual segmentation
(AVS)~\cite{zhou2022avs,zhou2023avss,guo2025audio,jin2026simtoken,zhou2025think,zhou2026audit,zhou2025mettle},
a widely studied task in the audio-visual
community~\cite{zhou2021positive,zhou2022cpsp,zhou2025towards,zhou2025clasp,zhou2024label,zhou2024advancing,zhou2025dense,zhao2025multimodal,li2024object,li2025patch,han2026face,mao2024tavgbench,zhou2026mtavg,li2026mtavg,shen2023fine}.
AVS uses sounds captured in the scene to segment the objects producing those
sounds~\cite{zhou2022avs}; it does not involve a spoken referring query. In
MeViS-Audio, by contrast, the audio conveys a spoken motion description, while
the referred object need not emit any sound. The task does not prescribe how
speech should be encoded: a system may ground the query directly from speech
features or first transcribe it into text. In practice, a common design first
applies ASR and then reuses text-conditioned grounding models. \method{}
follows this route, recovering the linguistic instruction before grounding its
motion semantics over time.

However, a transcription-based solution must handle three coupled sources of
uncertainty. First, pronunciation variation, rapid or indistinct speech,
background noise, and confusions between acoustically similar phonemes can
cause an ASR system to substitute or omit a word. The transcript may remain
fluent even when a high-value semantic slot---such as the target category,
count, direction, negation, temporal order, or interaction role---has been
corrupted. The resulting semantic error then propagates into prompt generation
and may change which object or trajectory is considered valid. Second,
multiple same-class objects can be visually indistinguishable in an individual
frame, while the decisive behavior occurs only briefly, before or after an
interaction, or under camera motion and occlusion. Correct grounding therefore
requires evidence from complete trajectories, including motion phase and
target--reference roles, rather than appearance alone. Third, some expressions
deliberately have no matching target. Forcing a prediction in such cases
selects a plausible but semantically invalid trajectory, whereas an overly
aggressive absence decision can suppress a genuine target. The system must
therefore support an empty mask sequence while retaining a conservative way to
reconsider uncertain absence decisions.

To manage these uncertainties, we introduce \method. This candidate-first
framework separates query interpretation, SAM3.1 trajectory proposal,
motion-aware ranking, presence gating, mask replacement, and empty-only
recovery. A confidence-aware compiler maps the ASR transcript to a structured
motion program while retaining the original transcription evidence and
uncertain hypotheses. SAM3.1~\cite{Carion2025SAM3} then produces an
over-complete trajectory pool, from which camera-compensated motion features
and our TRACE ranker construct a ranked SAM3.1 base prediction and fallbacks. A
frozen lexical presence gate either retains this result as control-positive
(target-present) or provisionally suppresses it. For a control-positive case,
an available full-expression-conditioned SaSaSa2VA (SaSa)
track~\cite{niu2025sasasa} replaces the SAM3.1 mask; TRACE does not compare
scores across the two backends, and their masks are not averaged. Only outputs
that remain empty enter GPT-assisted recovery, where query adjudication,
SaSaSa2VA regeneration, and mask arbitration may conservatively fill the empty
result. Together, these stages preserve uncertainty, exploit temporal evidence,
and allow abstention without assuming that transcription or instance-grounding
errors are fully eliminated. Our main contributions are:
\begin{itemize}
  \item a confidence-aware structured query compiler that retains raw ASR
  evidence and represents target identity, count, direction, temporal order,
  and interaction roles as explicit motion constraints;
  \item TRACE, a complete-trajectory ranker that combines learned compatibility
  with camera-compensated, phase-aware, and target--reference evidence while
  retaining fallback hypotheses;
  \item an asymmetric output policy that couples a frozen lexical presence
  gate with control-positive SaSaSa2VA replacement and GPT-verified, empty-only
  recovery, without allowing recovery to overwrite a nonempty prediction.
\end{itemize}

\section{Task Formulation}
\label{sec:task}

Given a video \(V=\{I_t\}_{t=1}^{T}\) and a spoken motion query \(A\), the goal is
to predict a binary mask for every frame,
\begin{equation}
  \mathcal{M}=\{m_t\}_{t=1}^{T}, \qquad
  m_t\in\{0,1\}^{H_t\times W_t}.
  \label{eq:prediction}
\end{equation}
The output may contain one or multiple referred objects. If the spoken query
has no valid referent, the correct prediction is an all-zero sequence,
\(m_t=\mathbf{0}\) for every \(t\).

For target-present expressions, the challenge reports the region similarity
\(\mathcal{J}\), boundary accuracy \(\mathcal{F}\), and their mean
\(\JF=(\mathcal{J}+\mathcal{F})/2\). No-target accuracy \(\Nacc\) measures how
often an absent query produces an empty prediction, while target accuracy
\(\Tacc\) measures how often a present query produces a nonempty prediction.
The final challenge score is~\cite{lsvos2026}
\begin{equation}
  \Final = \frac{\JF+\Nacc+\Tacc}{3}.
  \label{eq:final_metric}
\end{equation}
This metric discourages both forced grounding of absent queries and overly
aggressive suppression of valid targets.

\section{Method}
\label{sec:method}

\subsection{Overview}

\begin{figure}[t]
  \centering
  \includegraphics[width=\textwidth]{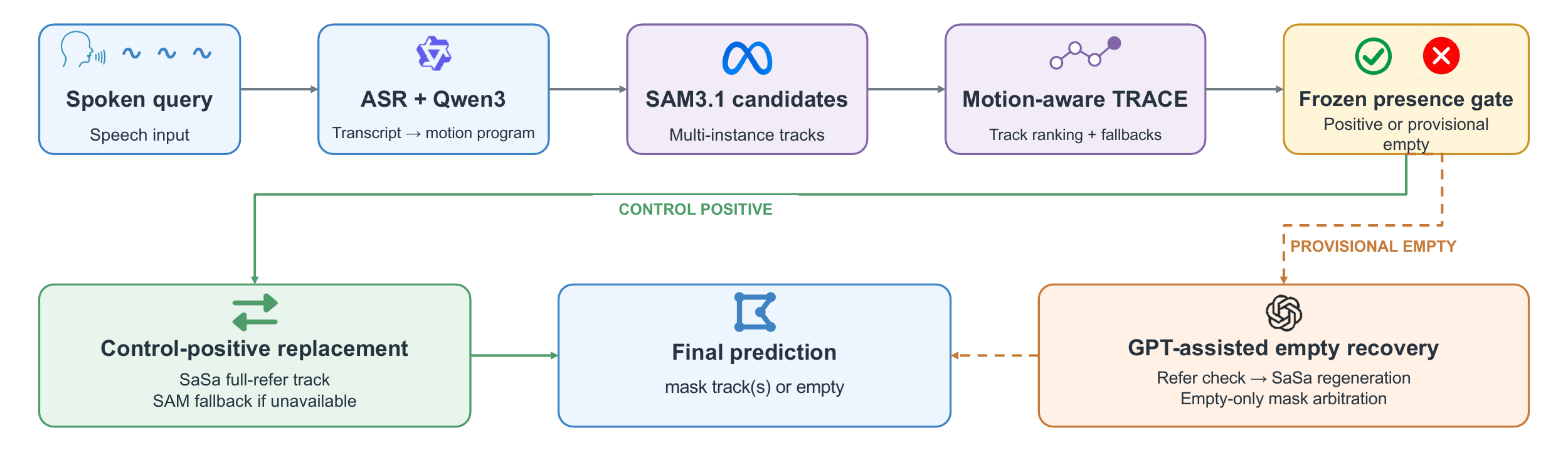}
  \caption{\textbf{Overview of \method.}
  A spoken motion query is transcribed and compiled into a structured motion
  program. SAM3.1 produces multi-instance tracks, which TRACE ranks using
  complete-trajectory evidence before a frozen lexical presence gate. For a
  control-positive result, an available full-expression-conditioned SaSaSa2VA
  track replaces the SAM3.1 mask; if that track is unavailable, the ranked
  SAM3.1 prediction is retained. Control-absent outputs bypass replacement. Any
  output that remains empty then enters GPT-assisted query adjudication,
  SaSaSa2VA regeneration, and mask arbitration. Recovery never overwrites a
  nonempty prediction.}
  \label{fig:method}
\end{figure}

\Cref{fig:method} summarizes \method. The framework delays commitment to a
single mask track until it has accumulated evidence over the complete video.
After ASR and structured query compilation, SAM3.1 enumerates entity-prompted
trajectories. TRACE ranks this SAM3.1 candidate pool using motion, relation, and
role evidence, after which a frozen lexical presence gate either retains the
ranked base prediction as control-positive or converts it into a provisional
empty output. The ranked SAM3.1 base prediction and its fallbacks are fixed
before the downstream SaSa stages; \emph{frozen} describes this decision boundary rather
than model training. For a control-positive result, an available
full-expression-conditioned SaSaSa2VA track replaces the SAM3.1 mask without
cross-backend score comparison. Only outputs that remain empty enter a separate
GPT-assisted path for query adjudication, SaSaSa2VA regeneration, and mask
arbitration. Recovery can fill an empty output but cannot modify a nonempty
prediction.

\subsection{Speech Transcription and Structured Query}

We first transcribe each spoken motion query \(A\) with the faster-whisper
implementation of Whisper large-v3~\cite{radford2023robust}. The raw transcript,
word-level timing, and confidence evidence are retained so that later
interpretation cannot silently overwrite the original query. A frozen Qwen3
instruction model~\cite{yang2025qwen} then compiles the transcript
\(q_{\mathrm{asr}}\) into a structured motion program
\begin{equation}
  z = f_{\mathrm{parse}}(q_{\mathrm{asr}}),
\end{equation}
where \(z\) contains a target category and count, segmentation prompts,
reference entities, spatial constraints, and a sequence of motion atoms. Each
motion atom records an action, direction, interaction reference, temporal
phase, and role of the target. Separating the target from reference entities
prevents an interacting object from being accidentally returned as the final
mask. The compiler records candidate ASR repairs and unresolved ambiguities,
but does not treat an uncertain repair as verified fact. This representation
therefore limits semantic error propagation without assuming that ASR errors
are fully corrected.

\subsection{Candidate Generation and Propagation}

The structured program is converted into separate target and reference prompts
for SAM3.1~\cite{Carion2025SAM3}. A candidate track is a complete sequence
\(c_k=\{m_{k,t}\}_{t=1}^{T}\), where an empty mask is used when object \(k\) is
not visible in frame \(t\).
For each prompt, SAM3.1 grounds one or more seed objects and propagates their
object states across the video, producing identity-consistent mask tracks. We
retain each track independently together with its
per-frame mask, bounding box, centroid, area, visibility, generator confidence,
and prompt role. These tracks form the candidate pool used to construct the
ranked SAM3.1 base prediction.

\subsection{Motion-Aware Candidate Ranking and Base Prediction}

\paragraph{Camera-compensated trajectory features.}
Motion expressions are evaluated on complete trajectories rather than isolated
frames. Global camera motion is estimated between neighboring frames with
sparse optical flow and robust affine fitting; phase correlation is used when
the affine estimate is unavailable. Let \(p_t\in\mathbb{R}^2\) be the observed
centroid of a candidate track in frame \(t\). Applying the estimated global
camera transform between frames \(t\) and \(t+1\) to \(p_t\) gives
\(\widehat{p}^{\mathrm{cam}}_{t+1}\), the position predicted under camera motion
alone. We define the camera-compensated residual as
\begin{equation}
  \Delta p^{\mathrm{res}}_t
  = p_{t+1}-\widehat{p}^{\mathrm{cam}}_{t+1}.
  \label{eq:camera_residual}
\end{equation}
The residual \(\Delta p^{\mathrm{res}}_t\) isolates candidate motion that
cannot be explained by global camera movement. Its direction and magnitude
therefore describe how the object itself moves. Over the complete track, we
summarize the residual sequence into candidate-level descriptors of motion
direction, strength, trajectory change, and early/middle/late activity. These
descriptors connect the structured motion program \(z\) to each candidate
track \(c\): for example, a rightward query is matched against horizontal
residual motion, while a late action is matched against the late-phase motion
summary. Together with visibility, area and shape variation, target--reference
geometry, and actor/patient role compatibility, these descriptors form the
candidate evidence used by the two scoring branches described next.

\paragraph{TRACE ranking and base prediction.}
We then introduce \textbf{Trajectory Ranking with Action-Conditioned Evidence
(TRACE)}, a compact learned scoring function that maps each query-conditioned
candidate feature vector to a scalar score. TRACE is learned from
candidate-level overlap supervision with video-grouped training. For a SAM3.1 candidate track \(c\) and
the structured motion program \(z\), let
\(S_{\mathrm{TRACE}}(c\mid z)\) denote the learned compatibility score. We
also aggregate the interpretable motion, direction, temporal-phase, spatial,
relation, role, visibility, and segmentation-confidence evidence into an
expert compatibility score \(S_{\mathrm{expert}}(c\mid z)\). The
camera-compensated descriptors above are included in the query-conditioned
TRACE input and also instantiate the motion, direction, and temporal-phase
components of the expert score. The two branches are combined as
\begin{equation}
  S(c\mid z)=\lambda S_{\mathrm{TRACE}}(c\mid z)
  +(1-\lambda)S_{\mathrm{expert}}(c\mid z).
  \label{eq:trace_score}
\end{equation}
where \(\lambda\in[0,1]\) is the fixed coefficient that balances the learned
ranker and expert evidence. A larger \(S(c\mid z)\) indicates that candidate
\(c\) is more compatible with the referred motion program. Candidates are
ranked in descending order of this score, and a count-aware rule may retain
multiple nonduplicate tracks for plural queries.

\paragraph{Frozen lexical presence gate.}
After candidate ranking, a frozen lexical presence gate decides whether the
ranked SAM3.1 base prediction should be emitted. The gate maps lexical evidence
in the compiled query to a presence score. A score below the fixed threshold converts
the result into a provisional empty output, denoted \emph{control-absent}; a
score at or above the threshold retains the selected track and ranked fallbacks
as \emph{control-positive}. The gate therefore suppresses plausible-looking
masks for likely no-target queries. A control-absent result bypasses the main
SaSa replacement, but the later GPT-assisted recovery path may reconsider its
provisional empty output.

\subsection{Control-Positive Replacement and Empty-Prediction Recovery}

\paragraph{Control-positive replacement.}
SaSaSa2VA~\cite{niu2025sasasa}, abbreviated as SaSa, jointly processes the
complete referring expression and temporally sampled video frames to decode one
or more dense mask tracks. This branch runs only after SAM3.1 ranking and the
frozen lexical presence gate. When the gate is control-positive and both the
ranked SAM3.1 base prediction and expression-conditioned SaSa track are
nonempty, the SaSa track directly replaces the SAM3.1 mask. TRACE does not compare scores
across the two backends, and their masks are not averaged. If SaSa produces no
usable track, the highest-ranked SAM3.1 track is retained. A control-absent
decision remains empty even if SaSa has produced a mask.

\paragraph{GPT query adjudication and SaSa regeneration.}
Only outputs that remain empty after presence gating and replacement enter
recovery. Here, GPT is first used as a query adjudicator rather than a mask
generator. It receives the ASR transcript and a chronological raw-video storyboard, repairs
an ASR interpretation only when supported by visual evidence, and returns a
normalized query, target-presence decision, constraint match, answer entity,
evidence frames, and confidence. SaSa is invoked a second time to generate a
recovery track only when GPT predicts \textsc{present}, reports a \textsc{full}
constraint match, provides a nonempty answer entity, and reaches the fixed
confidence threshold.

\paragraph{GPT mask arbitration.}
GPT is used again after SaSa regeneration, now as a mask arbiter. It compares
the unmodified storyboard with a second storyboard containing the candidate
mask overlay, and checks referent identity, class, count, attributes, action,
actor/patient role, relation, temporal evidence, and mask geometry. GPT never
produces mask pixels in either role. The recovery track is accepted only when
the arbiter verifies its identity and semantic constraints above the fixed
confidence threshold. Failure, insufficient evidence, or disagreement
preserves the empty prediction.

Let \(P_i\), \(S_i\), and \(R_i\) denote the gated SAM3.1 base prediction, the
main-branch SaSa replacement, and the recovery candidate for expression \(i\),
respectively. The final decision is
\begin{equation}
  B_i=\begin{cases}
    S_i, & P_i\neq\varnothing\ \land\ S_i\neq\varnothing,\\
    P_i, & \text{otherwise},
  \end{cases}
  \qquad
  M_i=\begin{cases}
    R_i, & B_i=\varnothing\ \land\ \operatorname{accept}(R_i),\\
    B_i, & \text{otherwise}.
  \end{cases}
  \label{eq:asymmetric_merge}
\end{equation}
Here, \(\operatorname{accept}(R_i)\) denotes successful GPT mask arbitration.
Thus, recovery can fill an empty prediction but can never overwrite a nonempty
mask track. This asymmetry balances target recall against no-target suppression.

\section{Experiments}
\label{sec:experiments}

\subsection{Implementation Details}

We use faster-whisper large-v3~\cite{radford2023robust} with English decoding,
beam size 5, temperature 0, word timestamps, and no cross-segment text
conditioning. The structured query compiler uses
Qwen3-30B-A3B-Instruct~\cite{yang2025qwen} with deterministic decoding.
SAM3.1~\cite{Carion2025SAM3} retains up to 16 objects per prompt at a proposal
threshold of 0.25. SaSaSa2VA~\cite{niu2025sasasa} uses the fixed 4B checkpoint
and 100 uniformly distributed temporal samples in both control-positive
replacement and empty-prediction recovery. TRACE ensembles five video-grouped
models with a learned-score weight of 0.65. The frozen lexical presence
threshold is 0.5, and recovery uses chronological 16-tile storyboards. We use
\texttt{gpt-5.6-terra} with low reasoning effort for query adjudication and
\texttt{gpt-5.6-sol} with medium reasoning effort for mask arbitration; both
recovery gates use a minimum confidence of 0.85. No test annotation or
reference mask is accessed during inference or selection.

\subsection{Official Leaderboard}

\Cref{tab:leaderboard} follows the official organizer notification. Entries
that appeared on the live Codabench page but were not part of the official
final ranking are omitted. The StopTheRoll entry, implemented with \method,
ranked second in the official final ranking.

\begin{table}[htbp]
  \caption{Official final ranking on the MeViS-Audio test set, as confirmed by
  the challenge organizers. Higher is better for all metrics.}
  \label{tab:leaderboard}
  \centering
  \small
  \setlength{\tabcolsep}{13pt}
  \resizebox{\linewidth}{!}{%
  \begin{tabular}{clccccc}
    \toprule
    Rank & Team & \(\JF\) & \(\mathcal{J}\) & \(\mathcal{F}\) & \(\Nacc\) / \(\Tacc\) & Final \\
    \midrule
    1 & zzjrr & 0.5952 & 0.5698 & 0.6205 & 0.7931 / 0.9205 & 0.7696 \\
    \rowcolor{bluewash}
    2 & StopTheRoll & 0.5374 & 0.5123 & 0.5626 & 0.7241 / 0.8554 & 0.7057 \\
    3 & rrrrty & 0.4587 & 0.4271 & 0.4903 & 0.6552 / 0.9518 & 0.6886 \\
    \bottomrule
  \end{tabular}
  }
\end{table}

\method{} achieves a final score of \(0.705662\) and a \(\JF\) score of
\(0.5374\). Relative to the third-ranked entry, \method{} improves \(\JF\) by
\(0.0787\) and \(\Nacc\) by \(0.0689\), while its \(\Tacc\) is \(0.0964\)
lower. The nonuniform metric profile is informative: \method{}
combines stronger mask quality and no-target handling than the third-ranked
entry, but target-present accuracy remains a clear source of error. The
\(0.0578\) \(\JF\) gap to the first-ranked method also leaves room to improve
both region overlap and boundary quality.

\subsection{Qualitative Results}

\begin{figure*}[t]
  \centering
  \includegraphics[width=\textwidth]{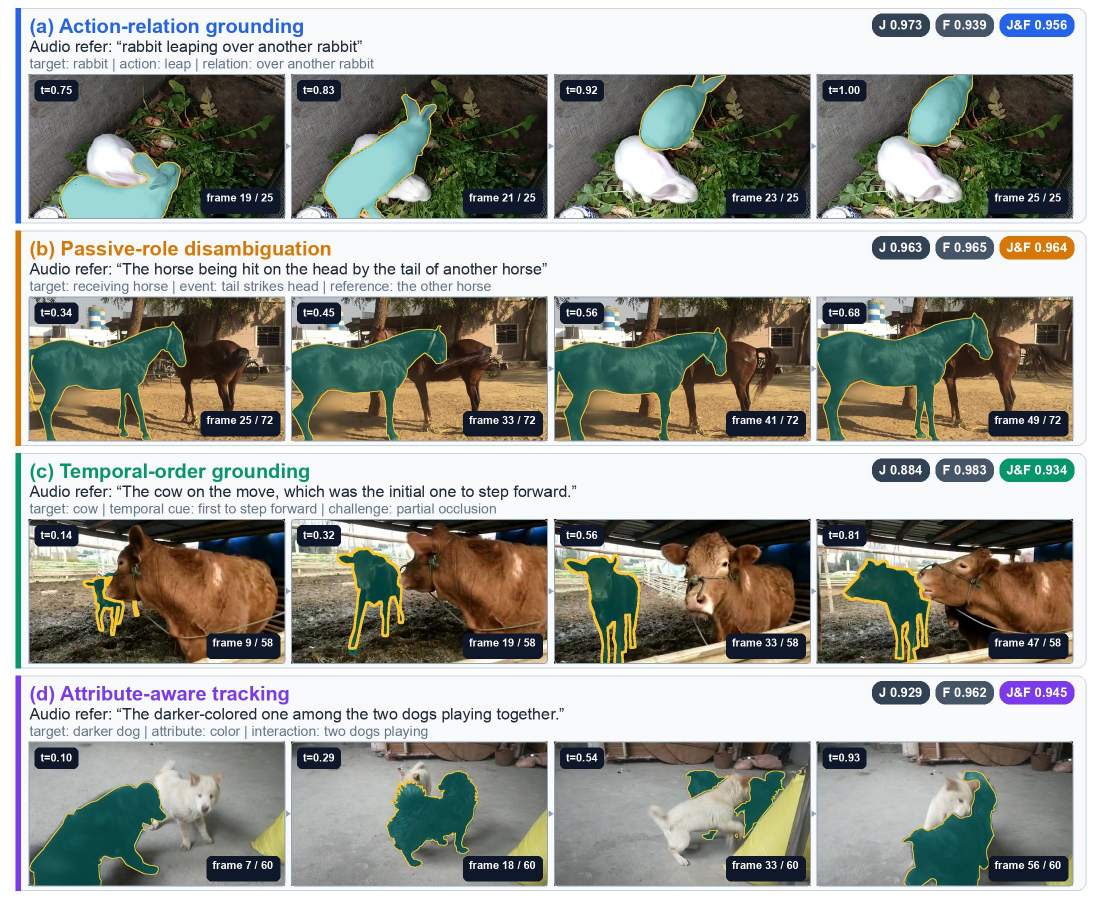}
  \caption{\textbf{Qualitative results on the MeViS-Audio validation set.}
  Teal fills and dark-teal contours denote our predictions, while yellow
  contours denote validation ground truth. The cases cover (a) a temporally
  sparse action relation, (b) passive-role disambiguation between
  same-category instances, (c) temporal-order grounding under partial
  occlusion, and (d) relative-attribute tracking through close interaction.
  The normalized timestamp \(t\) locates each frame in the complete video.
  }
  \label{fig:qualitative-results}
\end{figure*}

\Cref{fig:qualitative-results} presents four complementary cases from the
\texttt{valid\_u} split, chosen to span distinct grounding challenges. In
(a), \method{} follows the rabbit that performs a brief late leap over the
other rabbit, maintaining identity before, during, and after the interaction
\((\JF=0.9561)\). Case (b) requires selecting the patient of an event rather
than its most visibly active participant: the prediction consistently follows
the horse receiving the tail strike and obtains both region and contour scores
above \(0.96\). These examples support the use of temporally localized
relation and role evidence beyond static category matching, although they do
not establish robustness to every action or syntactic construction.

Case (c) exposes both a success and a residual failure mode. The target is the
cow that steps forward first, but it is heavily occluded early in the clip.
The displayed-frame overlap rises from \(0.7908\) in the early frame to above
\(0.92\) thereafter, while the sequence-level scores are
\(\mathcal{J}=0.8843\) and \(\mathcal{F}=0.9833\). Thus, the track preserves
identity as the cow emerges, but early occlusion still reduces region overlap.
In (d), \method{} follows the darker dog while two dogs overlap and exchange
relative position \((\JF=0.9453)\), illustrating attribute-aware identity
continuity during interaction. Taken together, these examples illustrate that
\method{} can handle temporally sparse actions, role ambiguity among
same-category instances, temporal-order grounding under partial occlusion, and
attribute-guided identity tracking during close interactions. They characterize
representative behavior but do not isolate the contribution of any individual
pipeline stage.

\section{Limitations}

The multi-stage design of \method{} remains vulnerable to error propagation.
Incorrect ASR or structured-query parsing can misdirect every downstream module, and TRACE
cannot recover an object that is absent from the SAM3.1 candidate pool. The
frozen lexical presence gate may also be brittle to unusual paraphrases or
ASR corruption. Although GPT-assisted recovery can reconsider an empty
prediction, its 16-frame storyboard may miss a brief discriminative event, and
the non-destructive policy cannot repair an incorrect but nonempty track. The
cow example in \Cref{fig:qualitative-results} further shows that severe early
occlusion can reduce region overlap even when later identity and boundary
tracking are accurate.
\method{} also requires several large models and multiple passes over each
video, increasing inference latency and implementation complexity. Moreover,
component-level ablations are unavailable for the final configuration, so the
individual gains from TRACE, SaSa replacement, and GPT-assisted recovery remain
unquantified.

\section{Conclusion}

We presented \method, a candidate-first framework for speech-guided referring
video object segmentation and the runner-up approach in the MeViS-Audio track
of the 8th LSVOS Challenge. \method{} converts speech into structured semantic and
temporal constraints, ranks SAM3.1 trajectories with camera-compensated motion
and relation evidence, replaces control-positive masks with
full-expression-conditioned SaSaSa2VA tracks, and invokes GPT-verified
SaSaSa2VA regeneration only for outputs that remain empty. The resulting entry
achieved an official final score of \(0.705662\). Qualitative cases illustrate
late-event grounding, passive-role disambiguation, temporal-order reasoning,
and attribute-aware identity preservation, while also exposing sensitivity to
early occlusion. These results suggest that delayed commitment is useful for
this task: query interpretation, trajectory ranking, presence gating, mask
replacement, and conservative recovery remain separate
until the final decision. Future work should retain this non-destructive
structure while reducing staged inference and improving robustness to noisy
speech and missing candidates.

\bibliographystyle{splncs04}
\bibliography{main}

\end{document}